\documentclass[pra,twocolumn,showpacs,superscriptaddress]{revtex4}
\usepackage{bm,graphicx,amsmath}

\begin{document}
\title{ Cavity QED quantum phase gates \\
for a single longitudinal mode of the intracavity field 
}

\date{\today}

\author{R. Garc\'{i}a--Maraver}
\affiliation{
Departament de F\'{i}sica,
Universitat Aut\`{o}noma de Barcelona,
E-08193, Bellaterra, Spain}
\author{R. Corbal\'{a}n}
\affiliation{
Departament de F\'{i}sica,
Universitat Aut\`{o}noma de Barcelona,
E-08193, Bellaterra, Spain}
\author{K. Eckert}
\affiliation{
Institut of Theoretical Physics, University of Hannover,
Appelstr.~2,
D-30167, Hannover, Germany}
\author{S. Rebi\'{c}}
\affiliation{
INFM and Department of Physics, Universit\`{a} di Camerino, 62032 Camerino, Italy}
\author{M. Artoni}
\affiliation{
Department of Chemistry and Physics of Materials, Via Valotti 9, 25133 Brescia, Italy }
\affiliation{
European Laboratory for Non-Linear Spectroscopy, \\
Via N. Carrara 1, 50019 Sesto Fiorentino, Italy}
\author{J. Mompart}
\email{jordi.mompart@uab.es}
\affiliation{
Departament de F\'{i}sica,
Universitat Aut\`{o}noma de Barcelona,
E-08193, Bellaterra, Spain}

\begin{abstract}
A single three-level atom driven by a longitudinal mode of a high-Q cavity is
used to implement two-qubit quantum phase gates for the intracavity field.
The two qubits are associated to the zero-and one-photon Fock states of
each of the two opposite circular polarization states of the field. 
The three-level atom yields the conditional phase gate provided the two polarization states 
and the atom interact in a $V$-type configuration and the two photon resonance condition is fulfilled. 
Microwave and optical implementations are discussed with 
gate fidelities being evaluated against several decoherence mechanisms  
such as atomic velocity fluctuations or the presence of a weak magnetic field. 
The use of coherent states for both polarization states is investigated to assess the
entanglement capability of the proposed quantum gates.
\end{abstract}

\pacs{42.50.Pq, 03.67.Mn, 32.80.-t}

\maketitle

\section{Introduction}
Along the last decade, cavity Quantum Electrodynamics (cQED) \cite
{Walther,Walther2,Feld,KimbleNLO,Orozco,Har1,Har2,Har3,Har4,Har5}
both in the microwave and optical regimes has been used to test
the most striking quantum features of single atoms interacting
with zero-, one-, or few-photon states. Some relevant examples are
the observation of photon trapping states and sub-Poissonian
statistics in the micromaser \cite{Walther}, the generation of
photon number states on demand \cite{Walther2}, the single atom
microlaser \cite{Feld}, nonlinear optics with single atoms and
photons \cite{KimbleNLO}, and non-classical statistics in
wave-particle quantum correlations \cite{Orozco}. In particular, a
series of seminal papers by S.~Haroche {\it et al}. \cite
{Har1,Har2,Har3,Har4,Har5} have shown that cQED with
long-lived atomic Rydberg states provides us with one of the most
simple systems to unambiguosly test the non-local nature of quantum
mechanics. In all these experiments, the Rabi oscillations between
a (vacuum) quantum field and an effective {\em two-level atom}
were used to entangle the cavity field with the atom. Moreover, the
passage of subsequent atoms through the microwave cavity with well
controlled velocities was used to entangle them via either real
\cite{Har1} or virtual photons \cite{Har5} allowing, therefore,
the creation of massive EPR pairs.

cQED devices hold great promise as basic tools for quantum
networks \cite{QNet} since they provide an interface between
computation and communication, i.e., between atoms and photons. In
this context, it is a very important task to look for techniques
to quantum engineer the state of the intracavity field. Very
recently it has been suggested the use of a {\em three-level atom}
in a cascade configuration \cite{Scully03} to entangle two
different longitudinal modes of the radiation field in one single
step. In particular, single and two bit quantum gates were
discussed with the number of photons ($n=0$ or $1$) of each mode being the
quantum bit of information. Although this proposal is very
interesting its eventual implementation presents two main drawbacks: 
(i) it requires a high-Q cavity that sustains two different longitudinal radiation modes;
and (ii) both modes must be adjusted to very particular
frequencies: one mode must be on-resonance with one of the bare
atomic transition frequencies while the second one must be tuned
to one of the dressed states built up by the first longitudinal
mode. We note here that three and multi-level atoms have been extensively investigated
in the past as a successful tool for many quantum optics applications \cite{nosaltres}
and, in particular, for quantum information \cite{KimbleNLO,Lukin,PhotonGun,Biswas,italians}.

In this paper, we propose a novel scheme that, while also using a three-level atom,
overcomes the disadvantages of the previously discussed proposal \cite{Scully03}. 
We will make use of a single longitudinal mode of the cavity to implement a
quantum phase gate (QPG) between the two qubits associated to the
zero and single photon states of the two opposite circular
polarization states of this mode. 
A precise control of the interaction time between the three-level atom and the mode will 
yield the conditional evolution needed to implement the QPG, 
provided that the atom and the two polarization
states interact in the so-called $V$-type configuration and that
the two-photon resonance condition is fulfilled.

The paper is organized as follows. In Sec.~II we will briefly review the interaction 
of a single three-level atom with a few photons field, discuss the basic ideas 
of the QPG implementation, and determine the explicit conditions for its realization.
In Section~III we will address some practical considerations for the physical 
implementation of the QPGs in both microwave and optical regimes. 
The application of the QPG to the intracavity field with both circular polarizations in a coherent state
will be discussed in Section IV as a method to entangle the intracavity field.
Finally, we will sumarize the proposal and present the main conclusions in Section~V. 

\section{Model}
The model we will use in this paper
is sketched in Fig.~1(a) and consists of a high-Q cavity with a
single longitudinal mode at angular frequency $\omega _{c}$ and a
three-level atom with its two allowed transition frequencies denoted by $%
\omega _{ac}$ and $\omega _{bc}$. These two atomic transitions
couple to the longitudinal cavity mode via the two opposite
circular polarizations $\sigma _{\pm }$ with coupling rates
$g_{\pm }$ and detunings $\Delta _{+}=\omega _{ac}-\omega _{c}$\
and $\Delta _{-}=\omega _{bc}-\omega _{c}$. For simplicity, in
what follows we will consider the completely symmetric case given
by $g_{+}=g_{-}\equiv g$ and $\Delta _{+}=\Delta _{-}\equiv
\Delta$. Later on, in Section III, we will address the case 
$\Delta _{+} \neq \Delta _{-}$ due to the presence of a 
stray magnetic field.

\begin{figure}[htb]
\centerline{
\hskip0.5cm
\includegraphics[width=8.5cm]{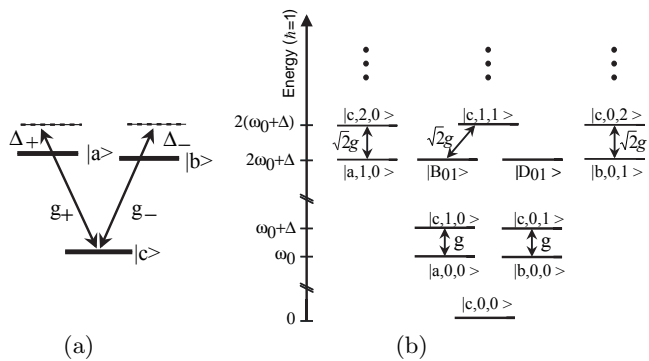}
} \centerline{ (a)\hskip4cm (b) \hskip1.7cm} 
\caption{(a) $V$-type
three-level configuration under investigation. Here $g_+$ ($g_-$)
is the vaccuum Rabi frequency of the coupling between the atom and
the right- (left-) hand circularly polarized field, and $\Delta_+$
($\Delta_-$) is the corresponding cavity-transition detuning. (b)
Dressed-states picture of (a) for $g_+= g_-$ $(\equiv g)$,
$\omega_{ac}=\omega_{bc}$ $(\equiv \omega_0)$, and $\Delta_+=
\Delta_-$ $(\equiv \Delta)$. $\left| B_{01}\right\rangle $ ($\left|
D_{01}\right\rangle $) is the bright (dark) state combination 
of $\left| a,0,1\right\rangle $ and 
$\left| b,1,0\right\rangle$ (see text). }
\label{fig:figs1ab}
\end{figure}

\subsection{Setting-up the idea}

It is very well known \cite{bright} that under the two-photon
resonance condition, the $V$-type system under investigation can
be appropriately described in the bright-dark states basis where
the ground state $\left| c\right\rangle $ couples only to a
particular combination of the atomic bare states $\left|
a\right\rangle $ and $\left| b\right\rangle $, namely the bright
state $\left| B\right\rangle $, while remaining uncoupled to the
orthogonal superposition, i.e., the dark state $\left|
D\right\rangle $. In this case, the three-level atom in
interaction with the two polarization modes becomes effectively
a two-level system, i.e., the atom exhibits Rabi oscillations
between the atomic ground state $\left| c\right\rangle $ and the
bright state $\left| B\right\rangle $. Thus, a single complete
Rabi oscillation takes the system back to the initial state with a
global phase that depends on the detuning. On resonance a $\pi $
phase is attained. 
These features of the $V$-type three-level system will be used later on
to quantum engineer the intracavity field.

In what follows we will define a qubit by the vacuum and single
photon Fock states. Thus, a single longitudinal cavity mode allows
to hold two qubits, one for the right-hand ($\sigma_+$) and one
for the left-hand ($\sigma_-$) circular polarization. 
To drive the conditional evolution between these two qubits, a
single three-level atom is initially prepared into the internal ground state
$\left| c\right\rangle $ and then the interaction is switched on for a controlled period of time. 
The particular mechanism to swith on/off the interaction will depend on the
physical implementation and, accordingly, will be discussed in the next Section.
Using the notation $\left| i\right\rangle \otimes \left| j\right\rangle
\otimes \left| k\right\rangle \equiv \left|
i,j,k\right\rangle $ where, respectively, $i$ denotes the atomic state while $j$ and $k$ the number of $%
\sigma _{+}$ and $\sigma _{-}$ polarized photons, the final state of the system after the interaction can 
be written, in general, as (see Fig.~1(b)):
\smallskip
\begin{eqnarray}
\text{Input\,State\,} & &\text{Output\,State}  \nonumber \\
\left| c,0,0\right\rangle &\rightarrow &\left| c,0,0\right\rangle  \nonumber
\\
\left| c,1,0\right\rangle &\rightarrow &c_{10}\left| c,1,0\right\rangle
+a_{00}\left| a,0,0\right\rangle \\
\left| c,0,1\right\rangle &\rightarrow &c_{01}\left| c,0,1\right\rangle
+b_{00}\left| b,0,0\right\rangle  \nonumber \\
\left| c,1,1\right\rangle &\rightarrow &c_{11}\left| c,1,1\right\rangle
+B_{01}\left| B_{01}\right\rangle , \nonumber
\end{eqnarray}
where the $c$'s, $a_{00}$, $b_{00}$, and $B_{01}$ are probability amplitudes whose
explicit value depends on the detuning and the interaction time.
For the symmetric case considered here
$c_{01}=c_{10}$ and $a_{00}=b_{00}$. 
$\left| B_{01}\right\rangle \equiv
\frac{1}{\sqrt{2}}\left( \left| a,0,1\right\rangle +\left|
b,1,0\right\rangle \right) $, ($\left| D_{01}\right\rangle \equiv
\frac{1}{\sqrt{2}}\left( \left| a,0,1\right\rangle -\left|
b,1,0\right\rangle \right) $) is the bright (dark) state
combination of $\left| a,0,1 \right\rangle $ and
$\left| b,1,0 \right\rangle $. 
For interaction times such that complete Rabi oscillations occur,
i.e., times for which the atom is brougth back to the internal state $\left| c \right\rangle$, only 
phases are left, whose explicit value depends on the number of oscillations and the cavity detuning. 
Therefore, looking for interaction times that yield complete Rabi oscillations both when the initial state is
$\left| c,1,0 \right\rangle $ (and, therefore, also for $\left| c,0,1 \right\rangle $) and when it is
$\left| c,1,1 \right\rangle $, two different QPGs can be implemented:
\begin{eqnarray}
\widehat{U}_{1} &=&e^{i\pi \delta _{j1}\delta _{k1}}\left| j,k\right\rangle
\left\langle j,k\right|  \label{gate1} \\
\widehat{U}_{2} &=&-e^{i\pi \delta _{j0}\delta _{k0}}\left| j,k\right\rangle
\left\langle j,k\right| . \label{gate2}
\end{eqnarray}
Each of these two QPGs together with arbitrary single qubit gates \cite{Scully03}
for both circular polarization states yields a universal set of quantum gates and,
therefore, these two QPGs can be used to entangle the two polarization states.
We want to note here that both $V$- and $\Lambda$-type three-level atomic configurations
are suitable for the proposal here discussed.
The additional advantatge of using a $V$-type scheme is that the common state 
$\left| c \right\rangle $
has lower energy than the other two atomic states 
and can be radiatively stable if it is the ground state.  

\subsection{Conditions for the gate operation}
To look for the conditions needed to implement (\ref{gate1}) and (\ref{gate2}) we start
by writing down the Hamiltonian of the system. In the rotating wave
approximation and the interaction picture, the truncated Hamiltonian of the
system restricted to the computational basis reads ($\hbar =1$):\
\begin{eqnarray}
H&=&g\left| a,0,0\right\rangle \left\langle c,1,0\right|
\ e^{-i\Delta t}
 +g\left| b,0,0\right\rangle \left\langle
c,0,1\right|\ e^{-i\Delta
t}  \nonumber \\
&&+ g\left| a,0,1\right\rangle \left\langle c,1,1\right|
\ e^{-i\Delta t} +g\left| b,1,0\right\rangle
\left\langle c,1,1\right| \ e^{-i\Delta
t} \nonumber \\
&&+ h.c. \nonumber \\
&=&g\ e^{-i\Delta t}\left| a,0,0\right\rangle
\left\langle c,1,0\right| +g\ e^{-i\Delta t}\left|
b,0,0\right\rangle \left\langle
c,0,1\right| \nonumber \\
&&+\sqrt{2}g\ e^{-i\Delta t}\left| B_{01}\right\rangle
\left\langle c,1,1\right| + h.c. 
\end{eqnarray}
where $g$ is the vacuum Rabi frequency. 
It will be shown in Section~III that the cavity decay and the spontaneous emission 
in modes other than the cavity mode can be neglected, based on a time-scale arguments, 
for a suitably chosen set of parameters. 
Accordingly, we now solve the Schr\"{o}dinger equation for this Hamiltonian
which, in our case, will provide the same information as the density matrix 
of the corresponding master equation.

It is clear from Eq.~(4) that we deal with three uncoupled two-level systems 
and a simple analytical solution can be obtained by integrating the
corresponding Schr\"{o}dinger equation. 
In each of these three cases, 
the probability amplitude of state $\left| c,j,k\right\rangle $ 
evolves in time according to
\begin{eqnarray}
c_{jk}(t) &\equiv &\left\langle c,j,k \right| \left. \psi (t)\right\rangle
\nonumber \\
&=&\frac{e^{i\Delta t/2}}{2}\Big{[} \left( 1-\frac{\Delta }{\Omega _{jk}}%
\right) e^{i\Omega _{jk}t/2}  \nonumber \\
& &  +\left( 1+\frac{\Delta }{\Omega _{jk}}\right) e^{-i\Omega _{jk}t/2}%
\Big{]} ,
\end{eqnarray}
with $\Omega _{01}=\Omega
_{10}=\sqrt{4g^{2}+\Delta ^{2}}$ and $\Omega
_{11}=\sqrt{8g^{2}+\Delta ^{2}}$, and where we have assumed
$c_{jk} (t=0) =1$ for $j,k = 0,1$. Note from (4) that
$c_{00} \equiv \left\langle c,0,0 \right| \left. \psi (t)\right\rangle $ 
does not evolve in time. Therefore, to implement the
first gate transformation (\ref {gate1}) one needs:
\begin{equation}
\frac{\Delta t}{2}=2\pi m;\quad \frac{\Omega _{01}t}{2}=2\pi n;\quad \frac{%
\Omega _{11}t}{2}=\left( 2p+1\right) \pi , \label{con0}
\end{equation}
where the integers $m$, $n$, and $p$ should fulfill the inequality
$2p+1>2n>2m\geq 0$. From the definition of $\Omega _{01}$ and
$\Omega _{11}$ it follows from Eq.~(\ref{con0}) that
\begin{equation}
\left( 2p+1\right) ^{2}=8n^{2}-4m^{2} , \label{con1}
\end{equation}
which is a Diophantine-type equation whereby for a fixed $n$ the
value of $m$ is determined through the detuning $\Delta $ according to the relation $\Delta /g=2m/\sqrt{%
n^2-m^2}$. The problem then reduces to find a $p$ closest to an
integer value that satisfies Eq.~(\ref{con1}). Similarly, the
implementation of the second quantum phase gate (\ref{gate2})
requires
\begin{equation}
\frac{\Delta t}{2}=2\pi m;\quad \frac{\Omega _{01}t}{2}=\left(
2n+1\right) \pi ;\quad \frac{\Omega _{11}t}{2}=\left( 2p+1\right)
\pi , \label{con3}
\end{equation}
leading to the equation
\begin{equation}
\left( 2p+1\right) ^{2}=2 \left( 2n+1\right) ^{2}-4m^{2} .
\label{con2}
\end{equation}
The corresponding inequality reads $2p+1>2n+1>2m\geq 0$, while
the relation between $m$, $n$ and the detuning is $\Delta
/g=2m/\sqrt{ \left( 2n+1\right)^2-m^2}$. We report in Table 1 the
best numerical solutions of (\ref{con1}) and (\ref{con2}) that, in addition,
minimize the interaction time $gt$. Note that the values for the detuning and the interaction time  
given in Table I are made dimensionless through the vacuum Rabi frequency $g$ which means
that these results are general in the sense thay they do not rely on any specific 
physical implementation. 

\bigskip
\begin{table}[htb]
\centerline{
\begin{tabular}{|c|c|c|c|c|c|c|c|c|c|}
\hline
$Gate$            & $m$  & $n$  & $p$      & $\Delta /g$ & $gt$    \\ \hline\hline
$\widehat{U}_{1}$ & $0$  & $6$  & $7.985$  & $0$         & $37.7$  \\
$\widehat{U}_{2}$ & $8$  & $10$ & $12.01$  & $2.353$     & $42.73$ \\
$\widehat{U}_{1}$ & $12$ & $15$ & $16.993$ & $2.667$     & $56.55$ \\
$\widehat{U}_{1}$ & $4$  & $12$ & $15.992$ & $0.707$     & $71.09$ \\
$\widehat{U}_{2}$ & $18$ & $21$ & $24.005$ & $3.062$     & $73.88$ \\
$\widehat{U}_{2}$ & $10$ & $15$ & $19.007$ & $1.689$     & $74.41$ \\
$\widehat{U}_{1}$ & $24$ & $28$ & $30.996$ & $3.328$     & $90.61$ \\
$\widehat{U}_{2}$ & $0$  & $14$ & $20.066$ & $0$         & $91.10$ \\
$\widehat{U}_{1}$ & $25$ & $29$ & $32.011$ & $3.402$     & $92.34$ \\
$\widehat{U}_{2}$ & $16$ & $22$ & $27.004$ & $2.022$     & $99.39$ \\ \hline
\end{tabular}
} 
\caption{Best numerical solutions to 
Eqs.~(\ref{con1}) and (\ref{con2}), sorted
by the required interaction time in dimensionless units.}
\end{table}
\bigskip

In order to check the validity of our proposal, we have
numerically integrated the Schr\"{o}dinger equation for the four
different input states in (1) and checked for both QPGs how much the output states
deviate in amplitude and phase from the exact
phase gate transformation. We have characterized the deviation by the 
following fidelity 
\begin{equation}
F=\left\langle \left| \sum_{j,k=0,1}
|c_{jk}^{out}|^{2}e^{\delta \phi _{jk}}\right| ^{2}\right\rangle ,
\end{equation}
where $c_{jk}^{out}\equiv \left\langle c,j,k\right| \left. \psi
_{jk}^{out}\right\rangle $, $\delta \phi _{jk}$ is the phase difference between
the phase acquired during the gate time and the exact phase of the gate defined in ($2$) and ($3$), 
and $\left\langle\dots \right\rangle$ denotes the average over the four different input
states. Figure 2 shows the results for
two different values of the cavity detuning: (a)
$\Delta = 0$, and (b) $\Delta = 2.35 g$.
Fidelities oscillate with peak values close to $F=1$ for the
particular interacting time values predicted in Table~I.   
\begin{figure}[htb]
\centerline{
\includegraphics[width=9cm]{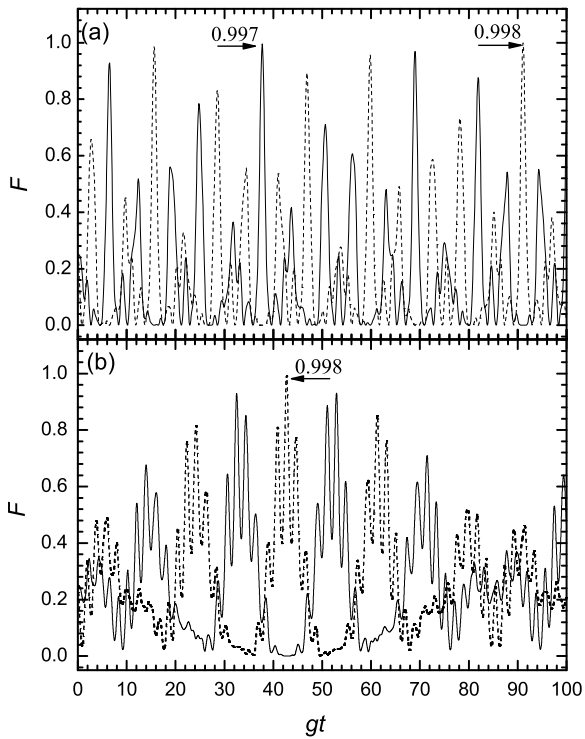}
} 
\caption{ Time evolution of the fidelity $F$ for both quantum phase
gates (solid curve for $\widehat{U}_{1}$ and dotted curve for
$\widehat{U}_{2}$). (a) $\Delta =0$, and (b) $\Delta =2.35 g$. }
\label{fig:fidelities}
\end{figure}

\section{Physical implementations}

Up to this point we have discussed the realization of the QPGs
for the intracavity field in general terms. 
Below we will give some practical considerations regarding the physical implementation
of the previous ideas in both microwave and optical regimes.

\subsection{Microwave regime}

In the microwave regime, 
single three-level Rydberg atoms crossing the high-Q cavity could be used to
implement the QPGs for the intracavity field. The interaction time can be 
controlled by an accurate selection of the velocity of the incident atoms.  
Typically, Rydberg atoms used in cQED yield vacuum-Rabi frequencies on the order of 
$g/2\pi \simeq 50$~kHz~\cite{Har1} which would imply 
gate times of a few tenths of a millisecond for the QPGs discussed in this paper.  
This time must be compared with the lifetime of a photon in high-Q microwave 
cavity that can be as large as few ms, and with the atomic
lifetime in Rydberg states that can be tens of ms \cite{Har1}.

We have investigated the robustness of the previously discussed QPGs 
against some of the experimental imperfections existing in the microwave regime. 
Figure 3 shows the fidelity for the implementation for the first gate in Table I
as a function of both the atomic velocity through the cavity and the intensity
of a uniform stray magnetic field along the cavity axis.
The magnetic field plays in general a
negative role as it breaks the degeneracy between atomic states
$\left| b \right\rangle $ and $\left| c \right\rangle $, i.e.,
yields $\Delta_+=-\Delta_- \neq 0$, so that the closed two-level
picture for the $\left| c,1,1 \right\rangle $ $\leftrightarrow$
$\left| B_{01} \right\rangle $ transition is no longer valid.
The relationship between the cavity detuning and the strength of the magnetic field
shown in Fig.~3 is given by
$\hbar \Delta_+=\mu_{B}g_{J}m_{J} B$
where $g_J$ is the gyromagnetic factor and $\mu_B$ is the Bohr magneton.
To be specific, we have choosen the following values $m_J=1$ and $g_J=3/2$ 
corresponding to $J=1$, $L=1$, and $S=1$.

Thus, gate realizations
with fidelities $F>0.99$ demand an accuracy of the atomic
velocities on the order of a few tenths of meters per second and
magnetic fields smaller than a few tenths of mG. Both requirements
can been achieved in present cQED experiments with Rydberg atoms \cite{Har5}.
Similar results are obtained for the rest of the QPGs shown in
Table I.
\begin{figure}[htb]
\centerline{
\includegraphics[width=9cm]{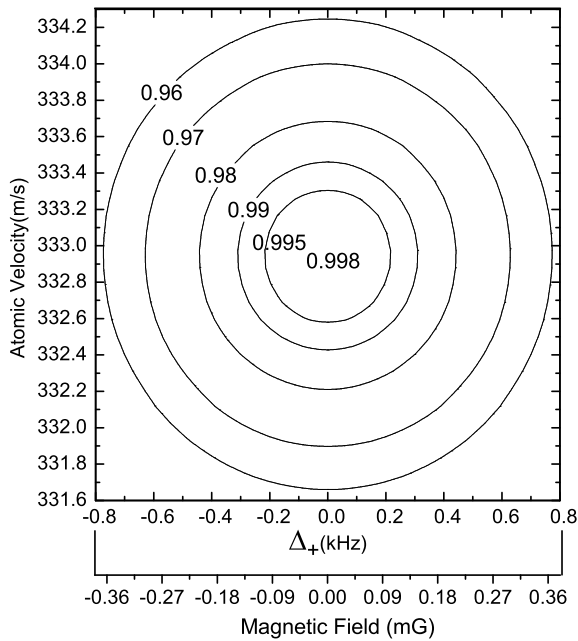}
}
\caption{Fidelity of the first $\widehat{U}_{1}$ gate
in Table I in the parameter plane atomic velocity versus
the intensity of a uniform magnetic field along the cavity axis.
The parameter setting is $\Delta = 0$, $g= 2 \pi \cdot 50$~kHz, $t=0.12$~ms, and
$L = 4$~cm being $L$ the effective cavity length.
}
\label{fig:velocity}
\end{figure}
\subsection{Optical regime}

By working in the optical regime, all the cryogenic complications encountered
in the microwave case needed to reduce the thermal photon noise can be avoided.
In the optical regime, a possible atomic candidate to drive the QPG in the intracavity 
field is Strontium \cite{strontium}. Depending on the
specific fine-structure component and on its four natural
isotopes (three of which are bosonic, $^{88} Sr$ ($82\%$), $^{86} Sr$
($10 \%$) and $^{84} Sr$ ($0.5 \%$)), a wide choice of transitions
with different $g$'s, linewidths and wavelengths are possible. The
inter-combination line $5 ^{1}S_0 - 5 ^{3}P_1$ of $^{88} Sr$, in particular,
spans two transitions that couple the ground state with vanishing
nuclear spin with two fairly long-lived ($\tau = 20$ $\mu$sec)
degenerate $J=1$ states. They both fall in the visible range
($\lambda = 689$ nm), and hence are easily accessible with common
semiconductor lasers; and both can acquire a large vacuum Rabi
frequency $g/2\pi \simeq 25$~MHz when high finesse micro-cavities
(F $\simeq 3 \times 10^6$) are used~\cite{rempe}. The coherent atom-field interaction needs 
to dominate over decoherence rates, hence the strong coupling regime of cQED is required. 
For the cavity decay rate $\kappa$ and spontaneous emission rate $\gamma$, 
this means that $g \gg (\kappa,\gamma)$. As the vacuum Rabi frequency $g$ is not constant throughout
the cavity mode volume, optimal results will be obtained for an atom trapped at the antinode of the cavity field. 
This is experimentally viable, as shown e.g. in Refs.~\cite{nacQED}, where trapping times up to 1 sec have 
been reported in the strong coupling regime. 

Results shown in Table I suggest that the gate operation can be
implemented over times $t_{gate}$ of the order of one microsecond.
Due to the very high finesse of micro-cavities~\cite{rempe}, 
lifetimes $\tau_{cav}$ of the order of few microseconds can possibly
be achieved, implying $t_{gate} \leq \tau_{cav}$, which is in
turn appreciably less than the atomic natural lifetime $\tau$. For the realization of gate fidelities $F>0.99$, a condition $t_{gate} \ll \tau_{cav}$ needs to be satisfied. This is feasible, given the rapid progress of optical micro-cavities over the recent years~\cite{rempe,nacQED}.

\section{Entangling the intracavity field}

Application of a two qubit QPG to an intracavity field with well
defined photon number clearly produces only a global phase. For
quantum computation purposes one also needs to produce single
qubit (local) operations for both polarizations. Methods for the
implementation of single qubit operations for the intracavity
field have been considered in \cite{Scully03}. 
To only demonstrate the entanglement capability, there is however 
a very simple and
accessible approach to entangle the two circular polarizations of
the intracavity field. 
We will not start with a well
defined photon number
in each mode, but initially inject into the hight-Q cavity
a product state consisting of a coherent state for both circular
polarizations:
\begin{equation}
| \alpha_+ , \alpha_-\rangle \equiv 
e^{-{{|\alpha_+ |^2 } \over 2} }
\sum_{n=0}^{\infty} {{\alpha_+^n } \over \sqrt{n!}} | n \rangle
\otimes
e^{-{{|\alpha_- |^2 } \over 2} }
\sum_{m=0}^{\infty} {{ \alpha_-^m} \over \sqrt{m!}} | m \rangle
\label{coh}
\end{equation}
Here
$|\alpha_+ |^2$ ($|\alpha_- |^2$) is the mean photon number of the
$\sigma_+$ ($\sigma_-$) polarization. 
This state is clearly separable. In Fig.~4 we use the concurrence
\cite{concurrence} $C \equiv 2 \left| c_{00}c_{11}-c_{01}c_{10}
\right|$ to quantify the entanglement between the two polarization states 
after the interaction with the atom as a function of the mean photon number of each polarization.
Notice that $C$ only takes into account the Hilbert space of zero and one photons, while other correlations
are ignored. The maximum concurrence, $C=0.73$, is reached when the mean photon number of both left and right
polarization is equal to $\left\langle N \right\rangle =0.5$.
\begin{figure}[htb]
\centerline{
\includegraphics[width=10cm]{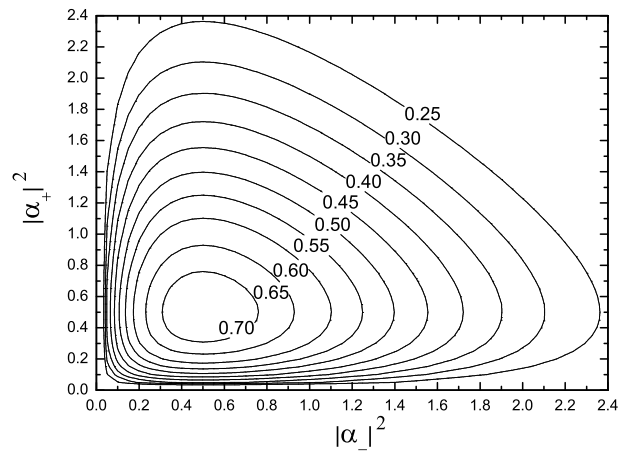}
}
\caption{
Concurrence between the two polarization states of the intracavity field (restricted to the computational basis)
as a function of the mean photon number of both polarizations.
Parameters correspond to the first $\widehat{U}_{1}$ gate sorted in Table~I. 
}
\label{fig:coherent}
\end{figure}

\section{Conclusions}
We have proposed a novel cQED technique to quantum enginneer the intracavity field. 
Single three-level atoms are used to implement two different QPGs between the vacuum and 
single photon states of the two opposite circular polarizations of a single longitudinal mode. 
QPG realizations with fidelities above 0.99 and gate 
times of around a tenths of a ms for the microwave regime and a few $\mu$s for the optical regime 
can be attained.
Some practical considerations such as the role of atomic velocity fluctuations or the presence of a uniform stray 
magnetic field along the cavity axis have been addressed, showing that the QPGs here discussed can be implemented 
with state of the art technology. 
We have analyzed a QPG realization where each circular polarization of the intracavity field 
is initially in a coherent state which, as it has been demonstrated, constitutes 
a simple method to entangle the two polarization states of the intracavity field.
We want to note that the ideas here discussed to quantum engineer the intracavity field could be 
extended to other cQED physical systems of current interest, e.g., solid state devices \cite{review},
such as superconducting electrical circuits, and SQUIDs \cite{Blais04,Yang04}.

Finally, we would like to note also that the use of three-level atoms in the optical regime 
has been previously discussed in the cQED literature \cite{KimbleNLO,PhotonGun}. 
For instance, single three-level atoms in a $\Lambda $-type configuration
have been used recently to deterministically produce a single-photon source \cite{PhotonGun}.

\begin{acknowledgments}
We gratefully acknowledge enlightening discussions with G.
Ferrari, J. Raimond, and D. Vitali. Financial support comes
from the MCyT (Spanish Government)
under contracts BFM2002-04369-C04-02 and HI2003-0075 (Acci\'{o}n Integrada Espa\~{n}a-Italia), 
from the DGR (Catalan Government) under contract 2001SGR00187,
and by the Deutsche Forschungsgemeinschaft through
the research program "Quanteninformationsverarbeitung", and by the Ministerio dell'Istruzione, dell'Universit\`{a}
e della Ricerca (MIUR) under contract IT1603 (Azione Integrada Italia-Spagna). 
R.~G.-M. acknowledges support from the MCyT for a grant (Ref. No.
BES-2003-1765)
\end{acknowledgments}

\end{document}